\documentclass[aps,prb,reprint,citeautoscript,superscriptaddress,showpacs]{revtex4-1}

\usepackage{graphicx,epsfig,float}
\usepackage{amssymb}
\usepackage{amsmath}
\usepackage[varg]{txfonts}

\allowdisplaybreaks
\begin{document}
\title{Atomic-scale defects restricting structural superlubricity: {\it Ab initio} study study
on the example of the twisted graphene bilayer}

\author{Alexander S. Minkin}
\email{amink@mail.ru}
\affiliation{National Research Centre ``Kurchatov Institute", Kurchatov Square 1, Moscow 123182, Russia}
\author{Irina V. Lebedeva}
\email{liv\_ira@hotmail.com}
\affiliation{CIC nanoGUNE BRTA, San Sebasti\'an 20018, Spain}
\affiliation{Catalan Institute of Nanoscience and Nanotechnology - ICN2, CSIC and BIST, Campus UAB, Bellaterra 08193, Spain}
\author{Andrey M. Popov}
\email{popov-isan@mail.ru}
\affiliation{Institute for Spectroscopy of Russian Academy of Sciences, Troitsk, Moscow 108840, Russia}
\author{Andrey A. Knizhnik}
\email{andrey.knizhnik@gmail.com}
\affiliation{Kintech Lab Ltd., 3rd Khoroshevskaya Street 12, Moscow 123298, Russia}

\begin{abstract}
The potential energy surface (PES) of interlayer interaction of twisted bilayer graphene with vacancies in one of the layers is investigated via density functional theory (DFT) calculations with van der Waals corrections. These calculations give a non-negligible magnitude of PES corrugation of 28 meV per vacancy and barriers for relative sliding of the layers of 7 -- 8 meV per vacancy for the moir\'e pattern with coprime indices (2,1) (twist angle 21.8$^\circ$). At the same time, using the semiempirical potential fitted to the DFT results, we confirm that twisted bilayer graphene without defects exhibits superlubricity for the same moir\'e pattern and the  magnitude of PES corrugation for the infinite bilayer is below the calculation accuracy. Our results imply that atomic-scale defects restrict the superlubricity of 2D layers and can determine static and dynamic tribological properties of these layers in a superlubric state. We also analyze computationally cheap approaches that can be used for modeling of tribological behavior of large-scale systems with defects. The adequacy of using state-of-the-art semiempirical potentials for interlayer interaction and approximations based on the first spatial Fourier harmonics for the description of interaction between graphene layers with defects is discussed. 
\end{abstract}
\maketitle

\section{Introduction}

Twisted graphene bilayer has attracted recently considerable attention due to its unique electronic properties such as the possibility to observe superconductivity \cite{Cao2018} and formation of a network of domain walls \cite{Lebedeva2019} with topologically protected helical states \cite{Vaezi2013,Zhang2013}. Relative rotation of graphene layers also gives rise to promising tribological properties \cite{Dienwiebel2004,Filippov2008,Androulidakisl2020,Liu2012,Vu2016},  namely structural superlubricity, i.e. the mode of relative motion of the layers with vanishing or nearly vanishing friction \cite{Hirano1990,Hirano1991}. This superlubric behavior can be used for elaboration of nanoelectromechanical systems based on electronic properties of graphene and relative sliding or rotation of graphene layers with respect to each other \cite{Poklonski13JCTN141,Popov2013a,Kang2015,Koren2015,Kang2016}. Rotation of graphene layers to incommensurate superlubric orientations is responsible for such phenomena as self-retraction of graphene layers \cite{Zheng2008,Vu2016,Popov2011a,Liu2012} and anomalous fast diffusion of a graphene flake on a graphite surface \cite{Lebedeva2010,Lebedeva2010a}. It should be mentioned that the phenomenon of structural superlubricity is observed not only for graphene-based systems \cite{Dienwiebel2004,Filippov2008,Androulidakisl2020,Liu2012,Vu2016} but also for multiwalled carbon nanotubes \cite{Zhang2013a}, graphene nanoribbons on gold surfaces \cite{Kawai2016},  graphene/hexagonal boron nitride heterostructure \cite{Song2018}, {\it etc.} (see review \cite{Hod2018} for more examples). For these 2D and 1D materials,  superlubricity is related with the incommensurate contact interface which is formed upon relative rotation of the layers of the same material to an incommensurate orientation or because of the lattice constant mismatch for heterostructures with layers of different materials.  In addition to graphene, a wide family of other 2D materials has been synthesized lately including hexagonal boron nitride (see Ref.~\onlinecite{Auwarter2019} for review), graphane \cite{Elias2009}, various transition metal dichalcogenides (see Ref.~\onlinecite{Shi2015} for review), phosphorene \cite{Churchill2014}, borophene \cite{Zhang2015}, germanene \cite{Yuhara2018}, {\it etc}. Heterostructures consisting of layers of different 2D materials should be also mentioned (see Ref.~\onlinecite{Geim2013} for review). Therefore, superlubricity can be expected for a wide set of incommensurate contact interfaces. 

Originally superlubricity for relative motion of 2D layers was discovered for nanoscale contacts between graphene flakes and graphite surface.\cite{Verhoeven2004,Dienwiebel2005,Filippov2008} To explain these experiments, a wide set of theoretical works and atomistic simulations were performed to study superlubricity between perfect rigid 2D layers and its loss via rotation of the layers with the same lattice constant to the commensurate ground state. \cite{Hirano1990,Verhoeven2004,Dienwiebel2005,Filippov2008,Bonelli2009,Guo2007,Shibuta2011,Xu2013,Wang2019} The calculations did not reveal any significant effect of atomic-scale defects on the static friction in the case of  incommensurate contacts between small graphene flakes and graphene layers \cite{Guo2007}. However, some decrease in the diffusion coefficient of a small graphene flake on a graphene layers was observed in simulations in the presence of defects and could be attributed to the increase in the dynamic friction force \cite{Lebedeva2010a}. It was shown also that superlubricity of very small flakes is restricted by pinning caused by distortions at the edges \cite{vanWijk2013,Mandelli2017}. This effect becomes negligible for large flakes \cite{Mandelli2017}. Recently not only nanoscale but also microscale and macroscale superlubricity between 2D layers \cite{Androulidakisl2020,Liu2012,Vu2016,Song2018} was observed. Moreover, robust superlubricity was achieved for systems with a lattice mismatch such as heterostructures \cite{Song2018} or similar layers under different tension applied \cite{Wang2019,Androulidakisl2020}. 
Theoretical studies \cite{Wijn2012,Muser2001} and recent experiments  \cite{Dietzel2013,Kawai2016} suggest that the superlubric friction force per unit area decreases with increasing the contact area. These observations generate interest in possible reasons which can restrict superlubricity of microscale and macroscale incommensurate contact interfaces.\cite{Liu2012,Mandelli2017,Hod2018,Koren2016} 

Up to now the following factors that restrict macroscopic robust superlubricity between 2D layers have been considered: 1) contribution of incomplete unit cells of the moir\'e pattern located at the rim area of the layer, \cite{Koren2016} 2) incomplete force cancellation within complete unit cells of the moir\'e pattern \cite{Koren2016} and 3) motion of domain walls in superstructures with large commensurate domains formed upon relaxation of  moir\'e patterns with spartial periods that are much greater than the domain wall width \cite{Liu2012,Mandelli2017,Hod2018}. Based on the studies of self-retraction motion of macroscopic graphene layers, it was suggested that the ultralow but nonzero friction for the layers with an incommensurate relative orientation can be induced by defects \cite{Liu2012} and the possibility of restriction of superlubricity by defects was discussed in the recent review devoted to superlubricity of 2D materials \cite{Hod2018}. 
The same argument was used to explain the nonzero (although very low) friction observed during macroscopic relative sliding of nanotube walls \cite{Zhang2013a}. Atomic-scale defects were shown to increase the dynamic friction in gigahertz oscillators based on relative sliding of nanotube walls \cite{Guo2005,Lebedeva2009} and to give the main contribution into the static friction during superlubric relative sliding and rotation of nanotube walls \cite{Belikov2004}. There is a lack of similar explicit studies of the influence of atomic-scale defects on friction in the case of macroscopic structural superlubricity for 2D materials. Here we perform {\it ab initio} calculations to investigate the effect of defects on 2D structural superlubricity by the example of twisted graphene bilayer with vacancies in one of the layers and discuss whether the defects can provide a dominant contribution to friction in the case of macroscopic superlubricity.

Registry-dependent semiempirical potentials \cite{Kolmogorov2000,Kolmogorov2005,Lebedeva2011,Lebedeva2012,Popov2012} were developed recently for description of interaction of graphene layers. They make possible modeling of relative sliding and rotation of the layers in large systems \cite{Lebedeva2010,Lebedeva2010a,Popov2011a}. However, these potentials were fitted to the results of {\it ab initio} calculations in the absence of defects. In the present paper, we consider the performance of one of these potentials, the Lebedeva potential \cite{Lebedeva2011,Lebedeva2012,Popov2012},  for interaction between the perfect graphene layer and the one with vacancies via comparison with the DFT results. Another approach for modeling of tribological behavior of a graphene flake on a graphene layer is based on approximation of the interaction energy between a single atom and a 2D hexagonal lattice by the first Fourier harmonics \cite{Verhoeven2004,Dienwiebel2005,Filippov2008}. The adequacy of such an approach in the case of the layers without defects was demonstrated not only for bilayer and few-layer graphene \cite{Ershova2010,Lebedeva2011,Popov2012,Lebedeva2012,Reguzzoni2012} but also for a variety of other 2D materials \cite{Lebedev2016,Lebedev2020,Jung2015,Kumar2015,Lebedev2017}.  Here we investigate whether this approach could still work in the presence of atomic-scale defects like vacancies.

The paper is organized in the following way. In Sec. II,
the model of the superlubric system and calculation methods are described. Sec. III is devoted to our results on structure and energetics of the vacancy, interlayer interaction for the moir\'e pattern with the perfect complete cell and influence of vacancies on the static friction. The possibility to use a simple approximation for the interaction between graphene layers with atomic-scale defects is also considered in Sec. III. The conclusions and discussion on tribological properties of macroscopic twisted graphene layers with numerous atomic-scale defects are presented in Sec. IV. 

\section{Methodology}
\subsection{Model of superlubric system}

An important characteristic which determines tribological properties of 2D materials is the potential energy surface (PES), that is the interlayer interaction energy as a function of the coordinates describing the relative in-plane displacement of the 2D layers. Particularly the PES determines directly  the static friction force for relative motion of the layers. To consider restriction of macroscopic superlubricity because of the presence of atomic defects, it is necessary to choose a model for atomistic calculations in which the contribution of perfect layers into the PES is negligible. To make such a choice here, we recall the results of atomistic modeling of superlubricity in the simple 1D case of double-walled carbon nanotubes. Namely, for double-walled nanotubes with commensurate walls at least
one of which is chiral, the PES of interwall interaction
is extremely flat and its corrugations are smaller
than the accuracy of calculations. This leads to the negligible static friction in the case of infinite walls (calculations using periodic boundary conditions) or finite walls with complete unit cells of the nanotube \cite{Kolmogorov2000,Damnjanovic2002,Vucovic2003,Belikov2004,Bichoutskaia2005}. For such nanotubes, due to only partial compatibility of helical symmetries of the walls, only very high Fourier harmonics of the interaction energy between an atom of one of the walls and the whole second wall contribute to the PES, whereas the contributions of other harmonics corresponding to different atoms of the nanotube unit cell are completely compensated \cite{Damnjanovic2002}. 
Thus, commensurate systems can exhibit superlubricity along with completely incommensurate systems (i.e. double-walled nanotubes with incommensurate walls). 
In the superlubric commensurate systems based on carbon nanotubes, edges \cite{Popov2013} and atomic-scale defects \cite{Belikov2004} are known to provide the main contribution into the static friction during relative sliding and rotation of the nanotube walls. 

Evidently infinite incommensurate systems without edges cannot be considered in the framework of DFT calculations with periodic boundary conditions. Incommensurate systems with edges make it difficult to study the restriction of superlubricity by defects since the contribution of edges to friction should be dominant for system sizes accessible to DFT calculations. Therefore, a superlubric commensurate system is a preferred choice for our DFT study. The results for double-walled nanotubes described above demonstrate that it is possible to use commensurate systems as models of superlubric systems in atomistic simulations. Here we show that it is also possible for 2D systems. 

Whereas twisted graphene bilayer is an incommensurate system in the general case, a set of commensurate orientations of the layers is observed for some special twist angles determined by coprime indices $(n,m)$ \cite{Xu2013,Campanera2007}. Only partial compatibility of translational symmetries of the layers in such commensurate moir\'e patterns is analogous to that for helical symmetries of the walls in double-walled nanotubes with commensurate walls at least one of which is chiral. In Section IIIB below, we confirm that  for the complete unit cell of the commensurate moir\'e pattern in absence of defects, contributions of individual atoms into the total PES are compensated within the calculation accuracy. Thus, only defects give rise to non-negligible PES corrugations and, correspondingly, to static and dynamic friction during relative motion of the layers. 

Therefore, in the present paper, we investigate the influence of defects on the static friction via the PES calculations for the complete unit cell of the commensurate moir\'e pattern. As an example, we consider graphene layers, one perfect layer and another with vacancies, rotated with respect to each other by  21.8$^\circ$ and forming the moir\'e pattern with coprime indices (2,1) (the commensurate moir\'e pattern with the smallest unit cell, Fig. \ref{fig:01}). Note that up to now the static friction between perfect twisted graphene layers has been studied only for motion of a finite smaller layer relative to the larger one for incommensurate relative orientations of the layers or the smaller layer including incomplete unit cells of the moir\'e pattern \cite{Hirano1990,Verhoeven2004,Dienwiebel2005,Filippov2008,Bonelli2009,Guo2007,Shibuta2011,
Xu2013,Wang2019,vanWijk2013,Mandelli2017}, that is only for the cases where the nearly total compensation of individual friction forces for atoms of the smaller layer is not possible. 

\subsection{Computational details}

\begin{figure}
   \centering
 \includegraphics[width=\columnwidth]{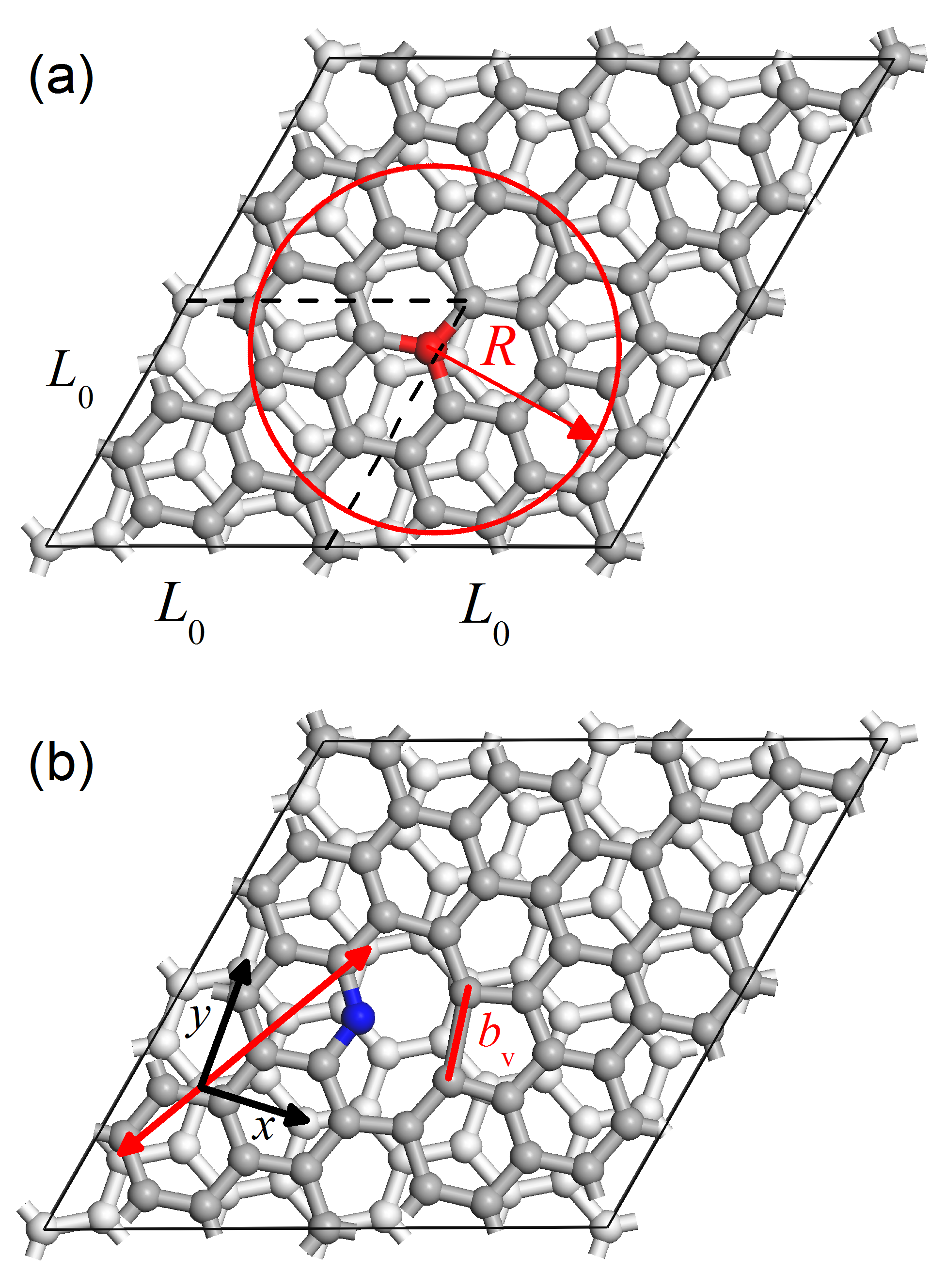}
   \caption{(Color online) $2\times2$ simulation cell of graphene layers rotated with respect to each other by 21.8$^\circ$ and forming the commensurate moir\'e pattern with coprime indices (2,1) in the absence of defects (a) and with a reconstructed vacancy (b). The upper and lower layers are coloured in dark and light gray, respectively. (a) The atom that is removed to create the vacancy is marked in red. The unit cell of the moir\'e pattern with sides $L_0$ is shown by the dashed lines. The circle of radius $R$ around the atom removed within which the contributions of atoms are taken into account in the approximation by the first Fourier harmonics is depicted. (b) The new bond of length $b_\mathrm{v}$ formed upon the vacancy reconstruction is shown by the red solid line. The atom with dangling bonds is coloured in blue. The axes $x$ and $y$ corresponding to Figs. \ref{fig:03} and \ref{fig:04} are included. The direction with small barriers for relative sliding of the layers according to the Lebedeva potential is indicated by the red double-headed arrow.} 
   \label{fig:01}
\end{figure}

The PES of twisted graphene bilayer was obtained through DFT and classical calculations. The spin-polarized DFT calculations were carried out with the VASP code \cite{Kresse1996}. The exchange-correlation functional of Perdew, Burke and Ernzerhof (PBE) \cite{Perdew1996} with the Grimme DFT-D2 dispersion correction \cite{Grimme2006} was applied. The parameters of the DFT-D2 correction optimized for bilayer graphene and graphite were used \cite{Lebedeva2016a, Leb_corr}. Interactions of valence and core electrons were described using the projector augmented-wave method (PAW) \cite{Kresse1999}. The Monkhorst-Pack  \cite{Monkhorst1976} method was applied for integration over the Brillouin zone. The maximum kinetic energy of plane waves was at least 500 eV. The Gaussian smearing of the width of 0.05 eV was used. The convergence threshold for self-consistent iterations was $10^{-9}$ eV. The bond length between carbon atoms in the perfect layer was taken equal to $l=$1.425~\AA, which is the optimal one for the PBE functional. Correspondingly, the trigonal unit cell of the moir\'e pattern with coprime indices (2,1) including 14 atoms in each perfect layer had equal sides of  $L_0=6.528$~\AA{} (Fig. \ref{fig:01}). The height of the simulation cell was 25~\AA. Periodic boundary conditions were applied.

First, the structure of the reconstructed vacancy in $2\times2$ and $3\times3$ simulation cells (which correspond to 4 and 9 unit cells of the moir\'e pattern and contain 56 and 126 atoms per the perfect layer, respectively) was studied. For that one atom was removed from the perfect graphene layer (Fig. \ref{fig:01}a), two of three two-coordinated atoms were brought closer to each other to form the bond giving rise to the 5/9 vacancy structure (Fig. \ref{fig:01}b) and geometry optimization was performed till the maximum residual force of 0.001 eV/\AA. The $10\times 10 \times 1$ k-point grid was used. The vacancy formation energy was calculated as  $\epsilon_\mathrm{v}=E_\mathrm{v}-\epsilon_\mathrm{gr}N_\mathrm{v}$, where $E_\mathrm{v}$ is the total energy of the system with the vacancy, $N_\mathrm{v}$ is the number of atoms in this system and  $\epsilon_\mathrm{gr}$ is the energy per atom in the perfect graphene layer.

To determine the optimal interlayer distance for the twisted layers, one unit cell of the moir\'e pattern with coprime indices (2,1) was considered. One of the layers was rigidly shifted perpendicular to the plane and the energy of the system was calculated as a function of the interlayer distance. The $14\times 14 \times 1$ k-point grid was used. The binding energy of the twisted layers per atom of the upper layer was found as $E_\mathrm{b}=(E_\mathrm{bi}-E_\mathrm{up}-E_\mathrm{low})/N_\mathrm{up}$, where $E_\mathrm{bi}$, $E_\mathrm{up}$ and $E_\mathrm{low}$ are the energies of the bilayer, upper and lower layers and  $N_\mathrm{up}$ is the number of atoms in the upper layer.

To compute the PES for the twisted layers, the $2\times2$ simulation cell of the moir\'e pattern  was considered for the upper layer with a single vacancy in the simulation cell and lower layer without defects (Fig. \ref{fig:01}b). The previously optimized structure of the layer with the reconstructed vacancy was used. The calculations were performed on the $14\times 14 \times 1$ k-point grid. The layers were placed at the optimal interlayer distance for the layers without defects and then the upper layer was rigidly shifted parallel to the plane with steps of 0.154~\AA{} and 0.130~\AA{} in the zigzag and armchair directions of the lower defect-free layer, respectively. 

The classical calculations were carried out using the registry-dependent Lebedeva potential \cite{Lebedeva2011,Lebedeva2012,Popov2012}. The parameters of the potential were fitted to the DFT data on the PES of co-aligned graphene layers. The calculations were performed with the parameters from Ref. \onlinecite{Popov2012}, cutoff radius of $R_c = 16.96$ \AA{} and height of the simulation box of 40 \AA. The structures of the layers were taken from the DFT calculations. The optimal interlayer distance was obtained for the $6\times6$ simulation cell of the moir\'e pattern of the twisted defect-free layers. To study the PES in the presence of vacancies, the upper layer with 9 equidistant vacancies in the $6\times6$ simulation cell (the $2\times2$ cell from the DFT calculations reproduced 3 times along each side of the simulation cell) placed at the optimal interlayer distance was rigidly shifted with respect to the lower defect-free layer with steps of  0.019~\AA{} and 0.016~\AA{} in the zigzag and armchair directions of the lower defect-free layer, respectively. The PES for the defect-free layers was calculated with the same steps in the $18\times18$ simulation cell of height 100 \AA{} for the cutoff radii $R_\mathrm{c}$ of the potential from 16 \AA{} to 50 \AA.

\section{Results}
\subsection{Structure and energetics of vacancy}

As known from the previous studies \cite{Wadey2016,Skowron2015,Ulman2014,Dai2011,Latham2013,Wu2013}, the energetically favourable structure of the vacancy in graphene corresponds to the 5/9 structure in which two of three atoms with dangling bonds form a new bond giving rise to 9- and 5-membered rings (Fig. \ref{fig:01}b). The comparison of the length $b_\mathrm{v}$ of the new bond and vacancy formation energy $\epsilon_\mathrm{v}$ with the data from literature is given in Table \ref{table:vac_struct}. A more detailed review of the previous results can be found in Ref. \onlinecite{Skowron2015}. It is seen from Table \ref{table:vac_struct} that the bond lengths $b_\mathrm{v}$ and vacancy formation energies $\epsilon_\mathrm{v}$ obtained here agree with the previously reported values lying in the ranges of 1.8--2.0 \AA{} and 7.4 -- 7.8 eV, respectively.  

Some of the previous calculations \cite{Dai2011}  predicted that the atom that is left with dangling bonds in the 5/9 structure (shown in blue in Fig. \ref{fig:01}b) exhibits significant deviation $d_\mathrm{v}$ perpendicular to the graphene plane, although the flat structure was observed in other papers \cite{Latham2013,Wadey2016} (Table \ref{table:vac_struct}). To clarify this, we considered several initial structures with out-of-plane deviation of the atom with dangling bonds of up to 0.4 \AA. However, we found that the final relaxed structure was always flat in the spin-polarized calculations and the flat vacancy structure was used in the further PES studies. It should be also noted that in the non-spin-polarized calculations, on contrary, the relaxed structures were characterized by significant out-of-plane deviations. Clearly, account of spin polarization related to the presence of an unpaired electron in the reconstructed vacancy is crutial for adequate description of the vacancy structure.

According to our calculations, the bond lengths $b_\mathrm{v}$ of the new bonds for the $2\times2$ and $3\times3$ simulation cells are different by 0.1 \AA{} and the vacancy formation energies $\epsilon_\mathrm{v}$ by  0.06 eV. These differences indicate that there is still some interaction of periodic images of the vacancies. However, they are sufficiently small to assume that the PES computed for the vacancy in the $2\times2$ simulation cell is close to that for the isolated vacancy. Note also that the differences in the bond lengths and vacancy formation energies for the two simulation cells considered are small compared to the scatter in the results of DFT calculations reported in literature (Table \ref{table:vac_struct} and Ref. \onlinecite{Skowron2015}).

\begin{table}
    \caption{Bond length $b_\mathrm{v}$ of the new bond in the 5/9 vacancy structure, vacancy formation energy $\epsilon_\mathrm{v}$ and out-of-plane deviation $d_\mathrm{v}$ of the atom with the dangling bonds in single-layer graphene obtained by spin-polarized PBE-DFT calculations here and in the previous papers for simulation cells with a different number of atoms $N$ (before vacancy formation).}
   \renewcommand{\arraystretch}{1.2}
   \setlength{\tabcolsep}{12pt}
    \resizebox{\columnwidth}{!}{
        \begin{tabular}{*{5}{c}}
\hline

Ref. & $N$  & $b_\mathrm{v}$   (\AA{}) & $d_\mathrm{v}$   (\AA{}) & $\epsilon_\mathrm{v}$ (eV)  \\\hline
This work & 56  & 2.077  & 0  & 7.70 \\\hline
This work & 126  & 1.977 & 0 & 7.64 \\\hline
\onlinecite{Latham2013} & 288  & 1.80 & 0 & 7.36 \\\hline
\onlinecite{Wadey2016} & 128  & 2.02 & 0 & 7.64 \\\hline
\onlinecite{Dai2011} & 72  & 1.95 & 0.184 & 7.67 \\\hline
\onlinecite{Ulman2014} & 56  &  &  & 7.72 \\\hline
\onlinecite{Wu2013} & 128  &  &  & 7.73 \\\hline
\end{tabular}
}
\label{table:vac_struct}
\end{table}

\subsection{Interlayer interaction for defect-free moir\'e pattern with complete unit cell}

The optimal interlayer distances $d_\mathrm{eq}$ and binding energies $E_\mathrm{b}$ for twisted graphene as well as their changes compared to the AB stacking ($\delta d_\mathrm{eq}$  and  $\delta E_\mathrm{b}$, respectively) obtianed here and the corresponding data available in literature are listed in Table \ref{table:energy}. Note that the PBE-D2 approach with the standard parameters for the dispersion correction used in Ref. \onlinecite{Ulman2014} underestimates the optimal interlayer distance and overestimates the binding energy of graphene layers \cite{Lebedeva2016a}. Using the parameters for the dispersion correction adjusted specifically for graphene \cite{Lebedeva2016a, Leb_corr}, we got more reasonable values of the optimal interlayer distance and binding energy  and slightly smaller changes in the interlayer distance and binding energy upon changing the twist angle from 0 to 21.8$^\circ$. Similar optimal interlayer distance and its change were reported previously in Ref. \onlinecite{Campanera2007}, although the variation in the binding energy obtained in that paper is smaller. The Lebedeva potential gives the changes in the interlayer distance and binding energy upon twisting the graphene layers closer to those from Ref. \onlinecite{Ulman2014} since it was fitted to the data obtained by the same PBE-D2 approach.

\begin{table}
    \caption{Calculated optimal interlayer distance $d_\mathrm{eq}$ (in \AA)  and binding energy $E_\mathrm{b}$ (in meV/atom) per atom of the upper layer for twisted bilayer graphene with the (2,1) moir\'e pattern and changes in the optimal interlayer distance $\delta d_\mathrm{eq}$ (in \AA) and binding energy $\delta E_\mathrm{b}$ (in meV/atom) as compared to the AB stacking. }
   \renewcommand{\arraystretch}{1.2}
   \setlength{\tabcolsep}{8pt}
    \resizebox{\columnwidth}{!}{
        \begin{tabular}{*{6}{c}}
\hline
 Ref. & Method &  $d_\mathrm{eq}$ & $E_{\mathrm{b}}$   & $\delta d_\mathrm{eq}$ & $\delta E_{\mathrm{b}}$    \\\hline
This work & DFT  & 3.40  & $-39.1$ & 0.08\footnote{See calculations within the same approach in Ref. \onlinecite{Lebedeva2016a}.}  &  3.6$^\mathrm{a}$\\\hline
\onlinecite{Campanera2007}\footnote{The results for graphite bulk.} & DFT  & 3.41  & & 0.09 & 2.7 \\\hline
\onlinecite{Ulman2014} & DFT  & 3.30 & $-48.0$ & 0.10 & 4.2 \\\hline
This work & \begin{tabular}{@{}c@{}} Lebedeva \\ potential \end{tabular}  & 3.46 &  $-41.8$ & 0.09 \footnote{See calculations within the same approach in Refs. \onlinecite{Lebedeva2011,Lebedeva2012}.}& 4.7 \\\hline
\end{tabular}
}
\label{table:energy}
\end{table}

\begin{figure*}
   \centering
 \includegraphics[width=\textwidth]{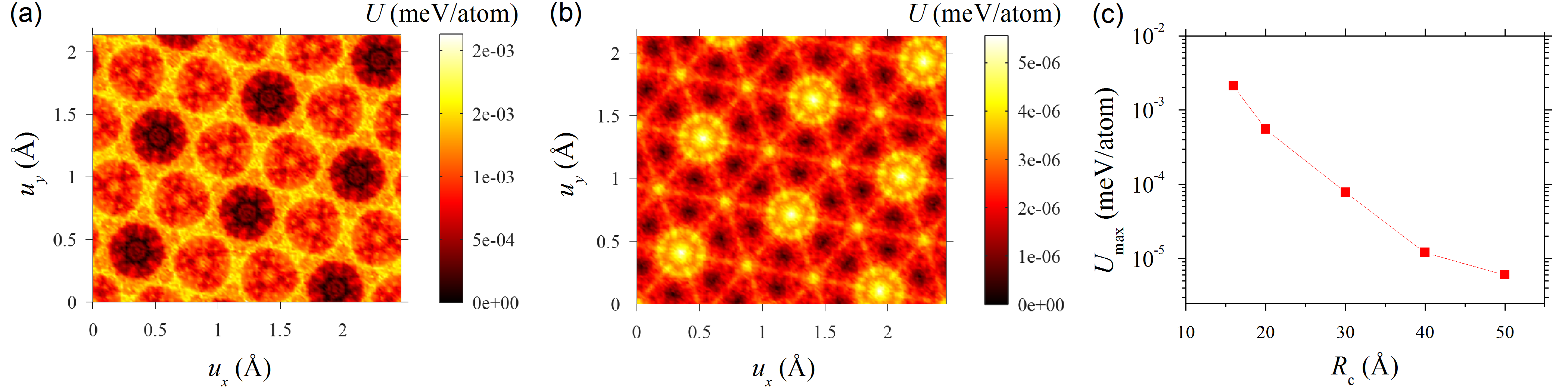}
   \caption{(Color online) Interaction energy $U$ (in meV per atom of the upper layer) of defect-free twisted graphene bilayer forming the moir\'e pattern with coprime indices (2,1)  as a function of relative displacement of the layers in the zigzag ($u_x$, in \AA) and armchair ($u_y$, in \AA) directions of the lower layer computed at the optimal interlayer distance of 3.46~\AA{} using the Lebedeva potential for interlayer interaction with the cutoff radii $R_\mathrm{c}$ of (a) 16 \AA{} and (b) 50 \AA{}. The energy is given relative to the minimum. (c) Calculated magnitude $U_\mathrm{max}$ of potential energy surface corrugation (in meV per atom of the upper layer) as a function of the cutoff radius $R_\mathrm{c}$ of the potential (in \AA).} 
   \label{fig:02}
\end{figure*}

To distinguish the effect of vacancies, we first computed the PES for defect-free twisted graphene layers forming the  (2,1) moir\'e pattern using the Lebedeva potential \cite{Lebedeva2011,Lebedeva2012,Popov2012}. It can be appreciated from Fig. \ref{fig:02} that for a finite cutoff radius $R_\mathrm{c}$ of the potential, the main contribution to this PES is provided by high spartial harmonics. Upon increasing the cutoff radius $R_\mathrm{c}$, the PES preserves its symmetry but the PES shape changes because of the cancellation of more and more harmonics. At the same time, the magnitude $U_\mathrm{max}$ of corrugation, i.e. the difference between the global energy maxima and minima at the same interlayer distance, decreases exponentially (Fig. \ref{fig:02}c). From the calculations with the cutoff radius of $R_\mathrm{c} = 50$ \AA{}, we estimate that the magnitude $U_\mathrm{max}$ of PES corrugation in the defect-free system does not exceed $6\cdot10^{-6}$ meV per atom of the upper layer. 

Different from our results, a considerable 
area contribution  into the force necessary for in-plane relative displacement of graphene layers (``area force" proportional to the number of complete unit cells) was found in Ref. \onlinecite{Koren2016} for twisted graphene bilayer with the same (2,1) moir\'e pattern as well as other moir\'e patterns with greater unit cells using the Kolmogorov-Crespi  potential \cite{Kolmogorov2005}. This discrepancy can be attributed to the following reasons. First, the cutoff radius used in Ref. \onlinecite{Koren2016} was only 16 \AA, i.e. 3 times smaller than the maximal cutoff radius used here. Thus, the area force found in Ref. \onlinecite{Koren2016} can be an artefact of the insufficient cutoff radius (see our Fig. \ref{fig:02}a for the same cutoff). Second, the shapes of the PES for infinite commensurate graphene bilayer without twist are different for the Lebedeva and Kolmogorov-Crespi  potentials. According to the DFT studies \cite{Ershova2010,Lebedeva2011,Popov2012,Lebedeva2012,Reguzzoni2012}, the PES of the commensurate graphene bilayer can be described well using only the first spatial Fourier harmonics (see also Subsection IIID). The parameters of the Lebedeva potential were specifically fitted to reproduce this property of the PES, whereas the shape of the PES for the Kolmogorov-Crespi  potential considerably deviates from that for the first spatial Fourier harmonics \cite{Lebedeva2012}. This means that amplitudes of higher harmonics of the PES for the interaction between an atom of one layer and the whole perfect adjacent layer for the Kolmogorov-Crespi  potential are considerably greater than those for the Lebedeva potential. This can in principle lead  to an incomplete cancellation of atomic contributions into the PES for moir\'e patterns with small unit cells even for the complete cell. Detailed studies of this problem are beyond the scope of the present paper. Moreover, the accuracy of DFT calculations of PESs for 2D materials and hence the accuracy of the calculations using potentials fitted to such PESs may be insufficient to consider effects related with high spatial Fourier harmonics. 

\subsection{Vacancy influence on static friction}

\begin{figure*}
   \centering
 \includegraphics[width=\textwidth]{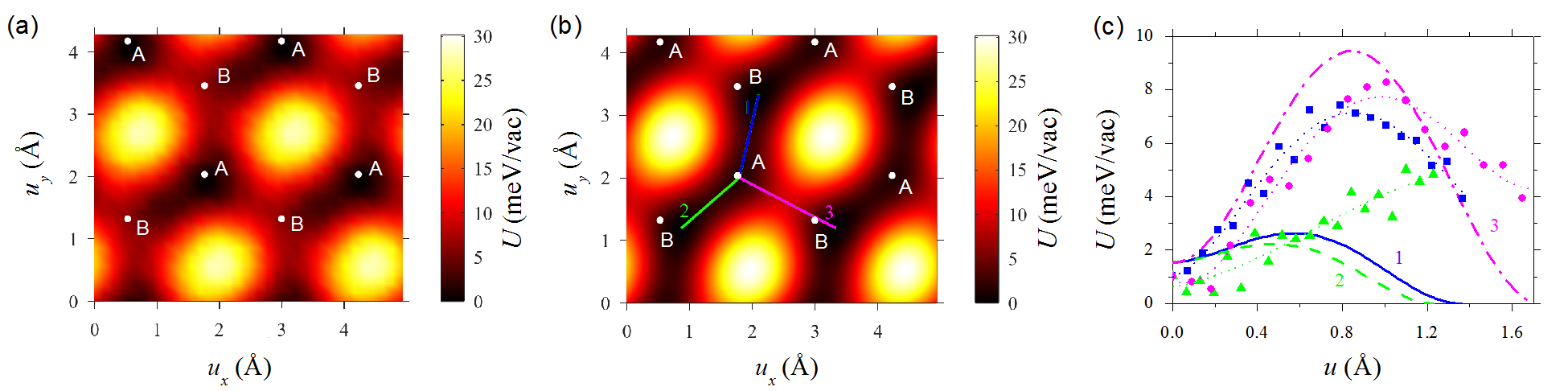}
   \caption{(Color online) Interaction energy $U$ (in meV per vacancy) of twisted graphene bilayer forming the moir\'e pattern with coprime indices (2,1) and one vacancy in the upper layer in the $2\times2$ simulation cell (see Fig. \ref{fig:01}) computed (a) via the DFT calculations and (b) using the Lebedeva potential for interlayer interaction as a function of relative displacement of the layers in the zigzag ($u_x$, in \AA) and armchair ($u_y$, in \AA) directions of the defect-free layer at the interlayer distances of (a) $d=$3.40~\AA{} and (b) $d=$3.46~\AA. The energy is given relative to the minimum. White circles correspond to relative positions of the layers where the atom of the upper layer removed to create the vacancy is on top of an atom of the lower layer. The sublattice A or B which the atom belongs to is indicated. (c) Interaction energy $U$ (in meV per vacancy) as a function of displacement ($u$, in \AA)  along the straight paths between adjacent energy minima of the potential energy surface computed with the Lebedeva potential, as depicted in panel (b). The results of the calculations with the potential are shown by lines and of the DFT calculations by symbols: 1 - blue solid line and squares, 2 - green dashed line and triangles, 3 - magenta dash-dotted line and circles. The dotted curves obtained by polynomial fitting of the DFT data are included to guide the eye.} 
   \label{fig:03}
\end{figure*}

When a vacancy is created in one of the layers, the magnitude $U_\mathrm{max}$ of corrugation becomes tens of meV per vacancy,  i.e. it is no longer negligible. The PES obtained by the DFT calculations for this system is shown in Fig. \ref{fig:03}a. The positions of minima and maxima on this PES are determined by the vacancy position with respect to atoms of the underlying layer. The maxima are displaced by 0.10 \AA{} from the stackings where the atom removed to create the vacancy (shown in red in Fig. \ref{fig:01}a) is located on top of centers of hexagons of the lower layer.  The positions of the minima are close to the stackings where the atom removed to create the vacancy is located on top of an atom of the lower layer. Only positions on top of atoms of one sublattice (we denote it A) correspond to the minima. The positions on top of atoms of the second sublattice (B) are neither minima, nor maxima. The minima are displaced from the on-top positions by 0.10 \AA. 

The magnitude $U_\mathrm{max}$ of PES corrugation according to the DFT calculations is 28.0 meV per vacancy (Fig. \ref{fig:03}a). It can be compared to the value for the coaligned commensurate bilayer (with zero twist angle), which is 15.6 meV per atom of the upper (adsorbed) layer according to the calculations with the same functional \cite{Lebedeva2016a}. The barriers for relative sliding of the layers between adjacent energy minima corresponding to the vacancy positions on top of atoms of the same sublattice A are 7--8 meV  (see also Fig. \ref{fig:03}c).  In the coaligned commensurate bilayer, the barrier is 1.7 meV per atom of the upper layer \cite{Lebedeva2016a}. 
For the twisted bilayer with the vacancy, the saddle points are located about 0.8 -- 1 \AA{} away from the minima. Therefore, we can estimate that the static friction force of about 12 -- 16 pN per vacancy should be applied to make the layers slide with respect to each other. For the coaligned commensurate bilayer, this force is about 6 pN per atom of the upper layer \cite{Popov2011,Lebedeva2016a}. It is clear that the magnitude of PES corrugation, barriers for relative sliding of the layers and static friction force in twisted bilayer are strongly reduced compared to the coaligned commensurate bilayer. For the simulation cell considered here that corresponds to the relatively high density of vacancies, the ratios of magnitude of PES corrugation, barriers for relative sliding of the layers and static friction force for the twisted and coaligned layers are only 0.03, 0.07--0.09 and 0.04--0.05, respectively. The net contribution of randomly distributed and orientated defects is discussed in the conclusion.

The PES computed using the Lebedeva potential is shown in Fig. \ref{fig:03}b. It has a number of similarities with the PES from the DFT calculations but also some differences. The maxima of the classical PES are located at exactly the same points as those of the \textit{ab initio} PES. There are shallow local minima at the same positions as the global minima on the \textit{ab initio} PES with the vacancy almost on top of the atoms of the A sublattice. However, new minima are also observed on the classical PES 0.36 \AA{} away from the positions where the vacancy is close to the atoms of the B sublattice and these minima are 1.6 meV lower in energy than the shallow ones. 

The magnitude $U_\mathrm{max}$ of PES corrugation for the Lebedeva potential is 30.2 meV (Fig. \ref{fig:03}b), which is only 8\% higher than the DFT value.  To estimate the barriers for relative sliding of the layers between energy minima, we considered the straight lines connecting the adjacent minima of the classical PES and computed the the energy variation along these lines (Fig. \ref{fig:03}c). The estimated barrier for one of these lines (path 3) agrees reasonably well with the DFT value,  9.4 meV vs. about 8 meV per vacancy, respectively. However, the barriers for sliding along the other lines are considerably smaller, 2.2 meV and 2.6 meV.
It can be indeed appreciated from Fig. \ref{fig:03}b that according to the semiempirical potential, there is a preferred direction for relative sliding of the layers in one of the zigzag directions. This direction is indicated by the red double-headed arrow in Fig. \ref{fig:01}b. This property is not observed in the DFT calculations that give similar barriers for different directions of motion (Figs. \ref{fig:03}a and c). Thus, the Lebedeva potential is able to describe the principal features to the PES of the twisted layers with a vacancy (symmetry, positions of the maxima and half of the minima, magnitude of corrugation, barriers across the preferred direction for sliding) but fails to describe its fine details (energies for relative positions of the layers with vacancies of top of atoms of the B sublattice, barriers along the preferred direction for sliding).

To investigate how the interaction between vacancies affects the PES, we also performed PES calculations for 4 equidistant vacancies in the $6\times6$ simulation cell using the Lebedeva potential. The structure of the layer with vacancies was taken from the DFT calculations for one vacancy in the $3\times3$ cell. These calculations revealed only minor changes in the PES as compared to the results for the $6\times6$ simulation cell with 9 vacancies, i.e. one vacancy per the $2\times2$ cell, discussed above. The magnitude of corrugation increased to 30.6 meV per vacancy, i.e. only by 2\%. The relative energy of shallow minima increased to 1.8 meV per vacancy, i.e. by 14\%. The barriers along paths 1, 2 and 3 in Fig. \ref{fig:03}c became 2.1, 2.8 and 11.2 meV per vacancy, i.e. changed by $-20$\%, $+25$\% and $+20$\%, respectively. These relative changes are not large given the accuracy of DFT calculations or calculations with the potential fitted to the DFT results. For example, the DFT data on the barrier for relative sliding of commensurate graphene layers reported in literature vary in the wide range from 0.5 meV/atom to 2.1 meV/atom (see Refs. \cite{Kolmogorov2005,Reguzzoni2012,Ershova2010,Lebedeva2011, Lebedeva2016a} and references therein). Note that more consistent values of 1.55--1.62 meV/atom are obtained when the interlayer distance is fixed at the experimental one \cite{Lebedeva2016a}. Still the experimental data on the width of domain walls \cite{Alden2013} and shear mode frequencies in bilayer and few-layer graphene \cite{Popov2012} suggest somewhat different barriers of 2.4 meV/atom and 1.7 meV/atom, respectively. Therefore, the typical error of calculations of barriers for relative sliding of van der Waals-bound layers within the DFT and DFT-based approaches 40\%. Given this accuracy, the calculations for one vacancy in the $2\times2$ cell look sufficient for qualitative and even quantitative description of the PES.

\subsection{Approximation by the first Fourier harmonics}

Another approach that can be used to model interlayer interaction in large-scale van der Waals systems is based on the PES approximation by the first spatial Fourier harmonics \cite{Ershova2010,Lebedeva2011,Popov2012,Lebedeva2012,Reguzzoni2012,Lebedev2016,Jung2015,Kumar2015,Lebedev2017}. Such an approach is even cheaper computationally than those based on  semiempirical potentials. The possibility to reproduce the PES obtained by the DFT calculations in this way was demonstrated not only for graphene bilayer \cite{Ershova2010,Lebedeva2011,Popov2012,Lebedeva2012,Reguzzoni2012}, but also for h-BN \cite{Lebedev2016} and hydrofluorinated graphene \cite{Lebedev2020} bilayers and graphene/h-BN heterostructure\cite{Jung2015,Kumar2015,Lebedev2017}. The hypothesis that the possibility of approximation of the PES by the first Fourier harmonics is a universal property of diverse 2D materials was proposed \cite{Lebedev2020}. Let us discuss whether the approximation by the first Fourier harmonics can still be used in the presence of defects.

The interaction energy between a single atom and a 2D hexagonal lattice is described by the first Fourier harmonics as \cite{Verhoeven2004}
\begin{equation} \label{eq_at0}
   U_\mathrm{hex} = 2U_1\left[2\cos(k_xu_x)\cos(k_yu_y) + \cos(2k_yu_y)\right] + const,
\end{equation}
where $x$ and $y$ axes are chosen in the zigzag and armchair directions, respectively, $k_x = 2\pi/\!\sqrt{3}l$, $k_y = 2\pi/3l$  ($l$ is the bond length), $\vec{u}$ is the relative position of the atom with respect to the lattice ($\vec{u} = 0$ corresponds to the case when the atom is located on top of one of the lattice atoms) and parameter $U_1$ depends on the interlayer distance. The interaction of a single atom with a 2D honeycomb lattice consisting of two hexagonal sublattices can then be written as 
\begin{equation} \label{eq_at}
\begin{split}
   U_\mathrm{hon} =&2U_1\bigg[2\cos\left(k_xu_x\right)\cos\left(k_yu_y+\frac{\pi}{3}\right)\\  - &\cos\left(2k_yu_y+\frac{2\pi}{3}\right)\bigg] + const.
\end{split}
\end{equation}

\begin{figure*}
   \centering
 \includegraphics[width=\textwidth]{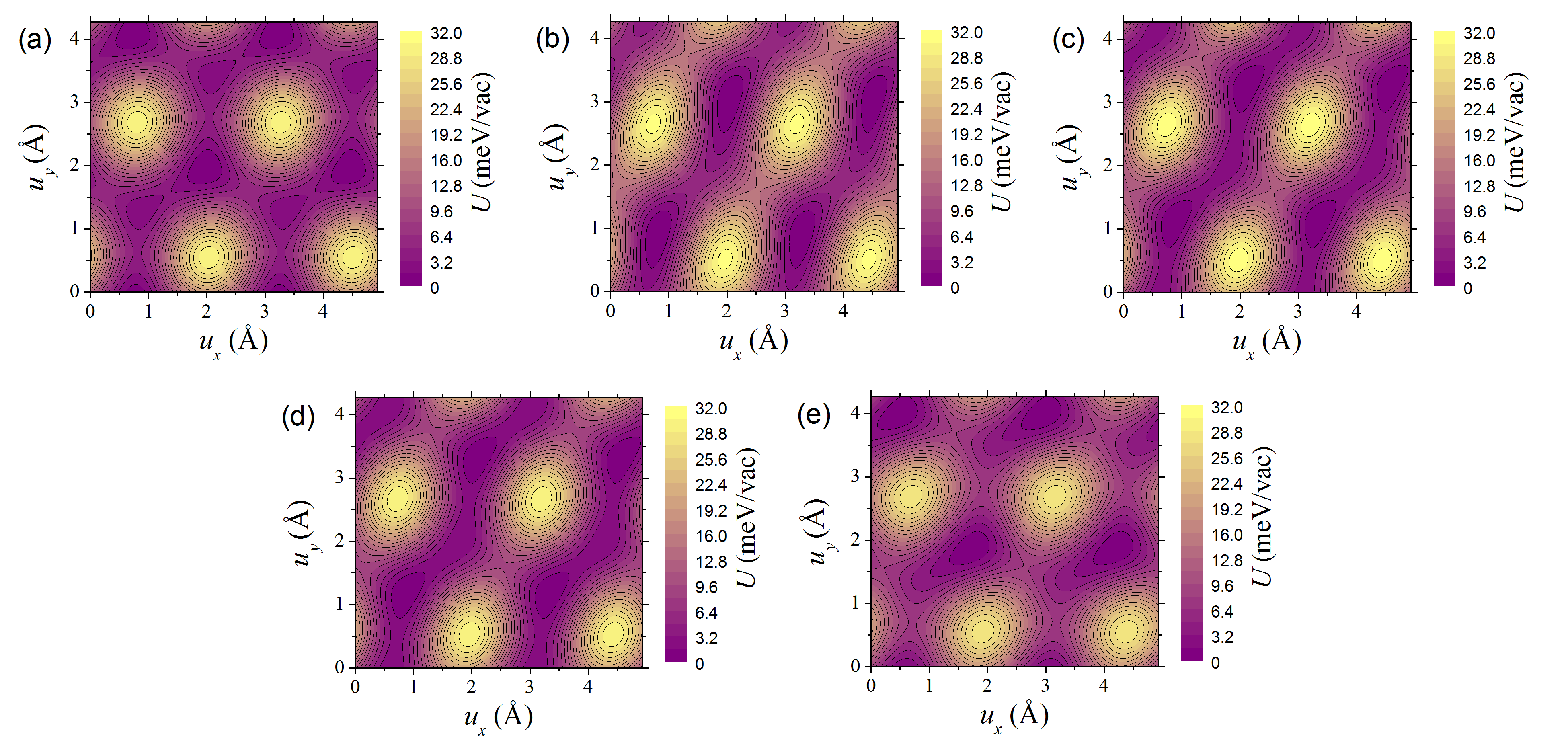}
   \caption{(Color online) Interaction energy $U$ (in meV per vacancy) of twisted graphene bilayer forming the moir\'e pattern with coprime indices (2,1) and one vacancy in the upper layer per the $2\times2$ simulation cell (see Fig. \ref{fig:01}) approximated by the first Fourier harmonics according to Eq. \ref{eq_at} as a function of relative displacement of the layers in the zigzag ($u_x$, in \AA) and armchair ($u_y$, in \AA) directions of the defect-free layer. The radius $R$ of the circle around the atom that is removed to create the vacancy within which the contributions of atoms are taken into account (see Fig. \ref{fig:01}a) equals: (a) 1.6  \AA, (b) 3.0 \AA{ }and (c) 5.0 \AA. All atoms of the simulation cell are taken into account in panels (d) and (e). The parameter of the approximation is $U_1=$2.7 meV. In panel (e), the parameter $U_1$ for the atom with dangling bonds in the structure with the vacancy (see Fig. \ref{fig:01}b) is reduced by 37\%. The energy is given relative to the minimum.} 
   \label{fig:04}
\end{figure*}

To investigate whether the PES of twisted graphene layers with defects can be described by the first Fourier harmonics, we summed up contributions corresponding to Eq. \ref{eq_at} for all atoms for the upper layer with the vacancy. From the classical calculations of the PES for co-aligned graphene layers at the interlayer distance of 3.463 \AA~optimal for the twisted defect-free layers, we estimated $U_1=2.7$ meV. The PES computed for the $2\times2$ simulation cell of the moir\'e pattern with coprime indices (2,1) with a vacancy based on Eq. \ref{eq_at} is shown in Fig. \ref{fig:04}d. As seen from comparison with Fig. \ref{fig:03}b, the shape and quantitative characteristics of the classical PES are well reproduced. The root-mean-square deviation of the approximation from the classical PES is only 0.05 meV per vacancy, which is within 0.2\% of the magnitude $U_\mathrm{max}$  of PES corrugation. The maximal deviation of the approximation is 0.13 meV, which is 0.4\% of $U_\mathrm{max}$, and $U_\mathrm{max}$ itself is different by only 0.2\%. 

For the twisted layers without defects forming an infinite commensurate moir\'e pattern, the contributions from all the atoms of one layer cancel each other (we checked this numerically for the (2,1) moir\'e pattern). Therefore, the PES can be also computed as a sum of differences $\Delta U_i  = U_{i,\mathrm{vac}}-U_{i,\mathrm{ideal}}$ of contributions $U_{i,\mathrm{vac}}$ and $U_{i,\mathrm{ideal}}$ given by Eq. \ref{eq_at} for atoms in the layer with the vacancy and the same layer before the vacancy formation. For the atom $i_0$ that is removed upon the vacancy formation (shown in red in Fig. \ref{fig:01}a), the first of these two terms is zero: $U_{i_0,\mathrm{vac}}=0$. It can be expected that $\Delta U_i$ goes to zero for atoms far from the atom removed, which have virtually the same position in the layers with and without the vacancy. Therefore, one can think on counting the contributions only from the atoms within some radius $R$ from the atom removed (Fig. \ref{fig:01}a). Our calculations show that already for the radius of $R=1.6$ \AA, which corresponds to account of only the nearest neighbours of the atom removed, the PES displays the preferred direction for sliding (Fig. \ref{fig:04}a). This feature becomes much more prominent upon inclusion of the second and further neighbours (Fig. \ref{fig:04}b). For $R=$5 \AA, which is close to the maximal possible for the $2\times2$ simulation cell considered, the PES looks already similar to the one computed using the Lebedeva potential (Fig. \ref{fig:03}b). For this radius, the root-mean-square deviation is 1 meV, which is about 4\% of the magnitude $U_\mathrm{max}$  of PES corrugation. The deviations of up to 2 meV, i.e. 7\% of $U_\mathrm{max}$ are observed. The magnitude of corrugation itself is higher by 5\%. This radius can be considered as a characteristic radius of the vacancy for the phenomena related to interlayer interaction.

The approximation by the Fourier harmonics can be also used to get insight into the origin of differences in the results of the classical and DFT calculations. In the classical calculations and approximation considered above, all the atoms of the upper layer interact in with the lower layer in the same way. However, atoms in the close vicinity of the vacancy carry different charges and spins and should interact in distinct ways. This can be taken into account by changing the parameter $U_1$ for different atoms. It can be expected that the contribution of the atom with dangling bonds differs the most (shown in blue in Fig. \ref{fig:01}b). Reducing for it the parameter $U_1$ by 20--50\%, the PES becomes qualitatively similar to the one obtained by the DFT calculations (Fig. \ref{fig:03}a). The former global minimum gets unstable and the barriers for relative displacement along and across the preferred direction for sliding become similar in magnitude (Fig. \ref{fig:04}e). The root-mean square deviation from the {\it ab initio} PES is minimized when $U_1$ for the atom with dangling bonds is reduced by 37\%. In this case, the magnitude $U_\mathrm{max}$  of PES corrugation for this approximation is only 0.4\% greater than the DFT result. The barriers for sliding are about 8 meV per vacancy, very close to the DFT estimate. The root-mean-square deviation from the \textit{ab initio} PES is 1.6 meV per vacancy, i.e. 6\% of the magnitude of PES corrugation. Probably this deviation can be further minimized by introducing slightly different $U_1$ for atoms forming the new bond in the reconstructed vacancy (Fig. \ref{fig:01}b). 

To summarize, the approximation by the first Fourier harmonics provides an extremely cheap alternative to DFT calculations and even calculations with classical potentials. Only the atoms in the close vicinity of the local defects, i.e. within 5 \AA{} in the case of the vacancy, need to be taken into account to get the reasonable accuracy. Differentiation of the parameters for atoms within the defect makes possible adequate approximation of DFT results even in the presence of spin-polarized defects.

\section{Discussion and conclusions}

The density functional theory with van der Waals correction 
was applied in the present paper to study the restriction of structural superlubricity coming from atomic-scale defects by the example of twisted bilayer graphene with coprime indices (2,1) of the commensurate moir\'e pattern and vacancies in one of the layers. For the purpose of this study, the vacancy structure and the PES of interlayer interaction for perfect twisted bilayer graphene were calculated. The structure of the isolated reconstructed vacancy was found to be flat in accordance with the majority of the previous studies \cite{Latham2013,Wadey2016} (see also Ref. \onlinecite{Skowron2015} for review). Corrugations of the PES for defect-free twisted graphene layers forming the  (2,1) moir\'e pattern were computed using the Lebedeva potential fitted to the DFT data and turned out to be less than the calculation accuracy. From the calculations with the largest cutoff radius of the potential considered, it can be concluded that the magnitude of PES corrugation in this case is less than $10^{-5}$ meV per atom of the upper layer. This contradicts the previous results \cite{Koren2016} obtained using the Kolmogorov-Crespi potential. However, the discrepancy should be mostly attributed to the low value of the cutoff radius of the potential used in that paper.

The DFT calculations for the twisted bilayer with vacancies in one of the layers gave the magnitude of PES corrugation of 28 meV per vacancy and the barriers for relative sliding of the layers in different directions of 7 -- 8 meV per vacancy. Thus, the presence of atomic-scale defects leads to non-negligible friction for twisted layers. For comparison, in the coaligned commensurate bilayer, the magnitude of PES corrugation
and barrier for relative sliding of the layers are 16 and 1.7 meV per atom of the upper
layer, respectively \cite{Lebedeva2016a}. Thus, the friction in the twisted bilayer with a reasonable density of vacancies is still small compared to the coaligned commensurate bilayer but large compared to defect-free twisted bilayer. 
 According to our DFT calculations, the static friction force of about 12 -- 16 pN per vacancy is required to induce sliding of the layers with respect to each other.  

Let us discuss applicability of the results obtained for the twisted graphene bilayer with the commensurate (2,1) moir\'e pattern to other 2D superlubric systems. As for the extreme PES flatness, previously the total compensation of atomic contributions into the PES (within the calculation accuracy) was demonstrated for complete unit cells of commensurate double-walled nanotubes with at least one chiral wall. \cite{Kolmogorov2000,Damnjanovic2002,Vucovic2003,Belikov2004,Bichoutskaia2005} For such nanotubes, the extremely flat PESs were found both via DFT calculations \cite{Bichoutskaia2005} and using various empirical potentials \cite{Kolmogorov2000,Damnjanovic2002,Vucovic2003,Belikov2004}. At the same time, the PES corrugations obtained by different calculation methods for the (5,5)@(10,10) nanotube with compatible symmetries of the walls differ by two orders of magnitude. \cite{Vucovic2003,Belikov2004,Bichoutskaia2005,Charlier1993,Palser1999,Saito2001,Bichoutskaia2006} Thus, the extreme flatness of the PES does not seem to be related with the nature of interlayer interaction but  rather with only partial compatibility of symmetries of the layers. We believe, therefore, that the compensation of atomic contributions to the PES in the considered case of twisted bilayer graphene is also a result of only partial compatibility of translational symmetries of the layers and similar cancellation can be expected for (2,1) moir\'e patterns of other 2D materials.
For commensurate double-walled nanotubes with at least one chiral layer, it was also shown that the smallest Fourier harmonics that contribute into the PES of the complete unit cell increase upon increasing the number of atoms in the unit cell, while the amplitudes of these harmonics decrease.\cite{Damnjanovic2002} Hence the PES corrugations for commensurate  moir\'e patterns with greater coprime indices should be even smaller than those for the considered moir\'e pattern with coprime indices (2,1).   

Furthermore, the extreme PES flatness for the complete unit cell without defects means that only defects (for example, vacancies) provide non-negligible contributions to the PES and can be considered as particles moving relative to the perfect layer. In such an imaginary picture, a change of the twist angle corresponds to the change in the orientation of these particles relative to the perfect layer. Evidently a small modification in the particle orientation leads to a small change of the PES. Thus, a small change of the twist angle from the commensurate  moir\'e pattern to  incommensurate one should lead to a small change in the contributions of defects to the total PES. Therefore, the model that we considered in the present study gives the results that are qualitatively valid also for incommensurate superlubric systems.

The PES calculations for the case of several disordered defects is beyond the scope of the present study. However, we would like to discuss briefly the influence of disordered defects on static and dynamic macroscopic friction in a superlubric system. In the case of random distribution and orientation of defects, their contributions into the static friction force cannot be summed up directly. The static friction force $F$ for a  disordered contact scales as $F \sim A^{1/2} \sim N_a^{1/2}$, where $A$ is the contact area and $N_a$ is the number of atoms in the contact. \cite{Muser2001,Dietzel2013} Similar scaling of the  friction force is expected for the contribution of disordered defects $F_d \sim A^{1/2} \sim N_d^{1/2}$, where $N_d$ is the number of defects. The edge friction force scales as $F_e \sim A_r^{1/2} \sim A^{1/4}$, where $A_r$ is the total rim area of incomplete cells of the moir\'e pattern \cite{Koren2016}. The static friction force between perfect incommensurate layers does not depend on the contact area \cite{Wijn2012}, that is the contribution of the perfect interface into the friction force $F_p \sim A^0$. Thus, the contribution of defects into the total static friction force $F=F_p+F_e+F_d$ between incommensurate 2D layers with disordered defects should become the dominant one when the macroscopic contact area $A$ is sufficiently large. Dissipation of the kinetic energy on  hills of the PES of interlayer interaction is the reason of dynamic friction during relative motion of 2D layers. \cite{Popov2011a} Therefore, a drastic increase of the PES corrugations due to the presence of defects should lead to the increase of dynamic friction in superlubric systems.

In the model system studied in our calculations, defects are present only in one layer. In the case of low densities of defects, the interaction between defects in the neighbour layers can be disregarded and defects from the both layers should contribute to the total static or dynamic friction in the same manner. Upon increasing the defect density, the interaction between defects in the neighbour layers (with probable formation of chemical bonds between the layers) should lead to further restriction of superlubricity. The presence of adjacent graphene layers or a substrate could also affect the PES. Nevertheless, previous DFT calculations \cite{Lebedeva2016a,Lebedeva2011,Popov2012} showed that differences in the PES corrugations and barriers for relative sliding of coaligned commensurate graphene layers in bilayer graphene and graphite are normally within 20\%, which is smaller than the scatter in the values of these physical quantities obtained using different DFT approaches \cite{Kolmogorov2005,Reguzzoni2012,Ershova2010,Lebedeva2011, Lebedeva2016a} and estimates of these quantities from the experimental measurements \cite{Alden2013,Popov2012}. Thus, we believe that the presence of additional 2D layers or a substrate for the superlubric system should not lead to a drastic change of contributions of defects into the PES of interlayer interaction in comparison with the bilayer system.

Let us now discuss the approaches that can be used for large-scale simulations of phenomena related to interlayer interaction in twisted bilayers with defects.
The magnitude of PES corrugation obtained using the semiempirical Lebedeva potential differs from the DFT result by only 8\%. Since the magnitude of PES corrugation determines dynamic friction related with dissipation of the kinetic energy of relative motion of the layers on such corrugations, the Lebedeva potential should be adequate for qualitative simulations of the  influence of atomic-scale defects on dynamic friction in systems with structural superlubricity. On the other hand, the semiempirical potential fails to describe regions of the PES around the minima and underestimates some of the barriers. Thus, it is not appropriate for static friction studies. This failure can be attributed to ignorance of spin-polarization effects.

An approximation based on the description of the interaction energy between atoms of one layer and the whole second layer via the first Fourier components provides an alternative to calculations with classical potentials for large systems and it is even cheaper computationally. 
Our calculations showed that such an approximation can reproduce closely the classical PES obtained using the semiempirical potential. Considering changes in contributions of atoms as compared to the defect-free bilayer, only the atoms in the close vicinity of the defect can be taken into account.  According to our calculations, it is sufficient to take into account atoms within 5 \AA{} from the vacancy to reproduce the classical PES with the error in the magnitude of PES corrugation of 5\% and root-mean-square deviation equal to 7\% of the magnitude of PES corrugation. This can be considered as an effective radius of vacancy defects for phenomena related with interlayer interaction. The approximation based on the first Fourier components can also reproduce the PES obtained by the DFT calculations once somewhat different parameters of the interaction are assumed for the atoms within the defect and in the perfect layer. For vacancy defects, the root-mean-square deviation of 6\% of the magnitude of PES corrugation is achieved when the amplitude of Fourier harmonics for the atom with dangling bonds is reduced by 37\% compared to the other atoms of the layer with vacancies.

The raw DFT data required to reproduce our findings are available to download from Ref.~\onlinecite{Minkin2021}.

\hfil
\section*{Acknowledgments}

A.A.K. and A.M.P. acknowledge support by the Russian Foundation for Basic Research (Grant No. 18-52-00002). I.V.L. acknowledges the European Union MaX Center of Excellence (EU-H2020 Grant No. 824143). The  work  was  carried  out using computing resources of the federal collective usage centre ``Complex for simulation and data processing for mega-science  facilities'' at  NRC  ``Kurchatov Institute'' (http://ckp.nrcki.ru).

The authors declare no conflict of interest.

\bibliography{prb2021-popov}
\end{document}